\documentclass[runningheads]{llncs}
\usepackage[T1]{fontenc}
\usepackage{adjustbox}
\usepackage{graphicx}
\usepackage{hyperref}
\usepackage{makecell}
\usepackage{caption}
\usepackage{changepage}
\usepackage{enumitem}
\usepackage{algorithmic}
\usepackage{algorithm}
\usepackage{amsmath, amssymb}
\usepackage[nolist]{acronym}
\usepackage{color}

\begin{document}
\title{Energy-Based Prior Latent Space Diffusion model for Reconstruction of Lumbar Vertebrae from Thick Slice MRI}
\titlerunning{Energy-based prior latent diffusion model for MRI vertebrae reconstruction}
%
\author{Yanke Wang\inst{1,*}\orcidID{0000-0003-1740-5269}\thanks{Corresponding author: Yanke Wang, \email{yankee.wann@gmail.com}.} \and
Yolanne Y. R. Lee\inst{2}\orcidID{0000-0001-6169-7065} \and
Aurelio Dolfini\inst{3} \and
Markus Reischl\inst{1}\orcidID{0000-0002-7780-6374} \and
Ender Konukoglu\inst{3}\orcidID{0000-0002-2542-3611} \and
Kyriakos Flouris\inst{3}\orcidID{0000-0001-7952-1922}}
\authorrunning{Y. Wang et al.}
%
\institute{Karlsruhe Institute of Technology, Hermann-von-Helmholtz-Platz 1, 76344 Eggenstein-Leopoldshafen, Germany \\
\email{yankee.wann@gmail.com, markus.reischl@kit.edu}
\and
Department of Computer Science, University College London, Gower Street, London WC1E 6BT, UK \\
\email{yolanne.lee.19@ucl.ac.uk}
\and
Department of Information Technology and Electrical Engineering, ETH Zürich, ETF E 111, Sternwartstrasse 7, 8092 Zürich, Switzerland\\
\email{adolfini@ethz.ch, \{ender.konukoglu,kflouris\}@vision.ee.ethz.ch}}

\maketitle              
\begin{abstract}
Lumbar spine problems are ubiquitous, motivating research into targeted imaging for treatment planning and guided interventions. While high resolution and high contrast CT has been the modality of choice, MRI can capture both bone and soft tissue without the ionizing radiation of CT albeit longer acquisition time. The critical tradeoff between contrast quality and acquisition time has motivated `thick slice MRI', which prioritises faster imaging with high in-plane resolution but variable contrast and low through-plane resolution. We investigate a recently developed post-acquisition pipeline which segments vertebrae from thick-slice acquisitions and uses a variational autoencoder to enhance quality after an initial 3D reconstruction. We instead propose a latent space diffusion energy-based prior
\footnote{The work is published in MICCAI Workshop on Deep Generative Models (DOI: \url{https://doi.org/10.1007/978-3-031-72744-3_3}), and the code is available at \url{https://github.com/Seven-year-promise/LSD_EBM_MRI}.}
to leverage diffusion models, which exhibit high-quality image generation. Crucially, we mitigate their high computational cost and low sample efficiency by learning an energy-based latent representation to perform the diffusion processes. Our resulting method outperforms existing approaches across metrics including Dice and VS scores, and more faithfully captures 3D features.

\keywords{MRI  \and Vertebrae \and Diffusion models \and Energy-based priors \and Image reconstruction.}
\end{abstract}

\begin{acronym}
\acrodef{mri}[MRI]{magnetic resonance imaging}
\acrodef{ct}[CT]{computed tomography}
\acrodef{srr}[SRR]{super-resolution reconstruction}
\acrodef{gan}[GAN]{generative adversarial network}
\acrodef{vae}[VAE]{variational autoencoder}
\acrodef{ddpm}[DDPM]{denoising diffusion probabilistic model}
\acrodef{dpm}[DPM]{diffusion probabilistic model}
\acrodef{lebm}[LEBM]{latent space energy-based model}
\acrodef{ebm}[EBM]{energy-based model}
\acrodef{mcmc}[MCMC]{Markov chain Monte Carlo}
\acrodef{lsdebm}[LSD-EBM]{latent space diffusion energy-based prior model}
\acrodef{dice}[DSC]{Dice's similarity coefficient}
\acrodef{vs}[VS]{volumetric similarity}
\acrodef{sen}[SEN]{sensitivity}
\acrodef{spec}[SPEC]{specificity}
\acrodef{nmi}[NMI]{normalized mutual information}
\acrodef{ck}[CK]{Cohen's kappa}

\end{acronym}

\section{Introduction}

Low back pain stands as the world's predominant musculoskeletal issue~\cite{wuGlobalLowBack2020}. For more serious symptoms, lumbar spine imaging and modeling is a critical tool used to aid in diagnoses and treatment planning. The lumbar spine is composed of five segments (L1-L5) and can exhibit significant variations~\cite{been2010new}, so patient specific models can provide valuable insight and inform possible treatment options. While \ac{ct} is particularly effective at capturing skeletal structures with high resolution and high contrast, it uses ionizing radiation and fails to capture soft tissue. Alternatively, \ac{mri} captures not only the vertebrae but also the disc spaces, spinal canal, and nerve roots without ionizing radiation but at the cost of acquisition time~\cite{bajger2021lumbar}.

One of the key factors of \ac{mri} for acquisition times is the slice thickness~\cite{laakso1997mri}.
While using thinner slices would improve the through-plane resolution, it greatly increases acquisition time
~\cite{sui2022scan}, which leads to patient discomfort and increased motion artifacts.
As a result, so-called `thick slice \ac{mri}' is typically used in clinical practice, prioritizing high in-plane resolution and faster acquisition times at the cost of through-plane resolution. Machine learning-based reconstruction can potentially recover missing details and allow detailed anatomical modeling by increasing the through-plane resolution while faithfully reconstructing fine details.


\begin{figure}[htb]
\centering
    \includegraphics[width=\textwidth]{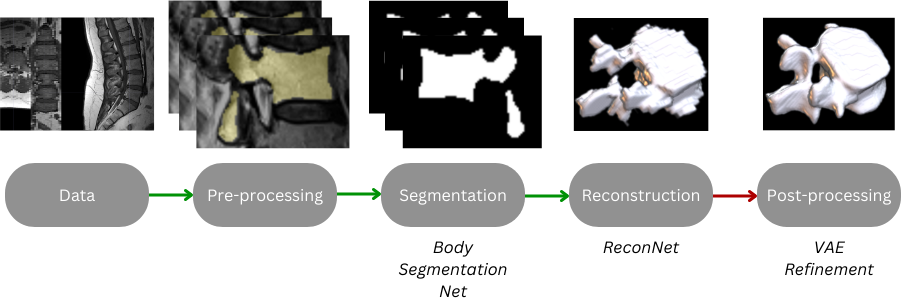}
    \caption{Schematic diagram of the segmentation and reconstruction of high-quality  lumbar vertebrae \ac{mri} images, with the proposed pipeline of ~\cite{turella2021high} shown in italics. We focus on the generative method of the post-processing step marked in red.} 
\label{fig:pipeline}
\end{figure}

A complete pipeline was introduced in \cite{turella2021high} (in italics, Fig.~\ref{fig:pipeline}) that segments \ac{mri} data into vertebral body masks, then turns these low-quality masks into high-quality \ac{ct}-like segmentations of the full vertebrae via their ReconNet, and finally refines the resulting 3D model via a \ac{vae}. ReconNet, a U-net based architecture, is trained on segmentations from widely available \ac{ct} lumbar spine datasets to 
generate highly detailed segmentations from distorted segmentations predicted from thick-slice acquisitions.
The model in \cite{turella2021high} uses the \ac{vae} as a post-processing step which takes a reconstruction from the ReconNet masks and outputs a more anatomically feasible reconstruction. However, the results of this automated pipeline are too ``smooth'' in comparison to the baseline 3D \ac{ct} reconstruction, lacking fine detail in the anatomy. More powerful anatomical priors have the potential to improve this last step of the pipeline. 


We propose the \ac{lsdebm} for enhancing 3D \ac{mri} reconstructions from refined segmentations of the lumbar spine. We aim for an expressive generative model capable of learning anatomically feasible structures while additionally retaining sharp individual sample details, which could be used for more accurate patient modeling in clinical practices. We investigate probabilistic generative models in effort to restrict the space of generated segmentations to the distribution of real segmentations. To this end, our model leverages the capabilities of diffusion models and energy priors while keeping computational costs manageable. 
The model is trained on high-quality vertebrae segmentations extracted from \ac{ct} images, in order to learn a prior on vertebrae structure that can be used to generate missing details of a given refined segmentation based on thick slice \ac{mri}, for example, from the output of ReconNet. 

Our contributions are the following: we propose a novel \ac{lsdebm} framework for image generation using an advanced energy-based latent for the diffusion model. We implement our model in the ReconNet pipeline, providing an updated easy-to-use tool. We evaluate its performance by testing the pipeline end-to-end with the modified final \ac{lsdebm} step using multiple evaluation metrics for a more detailed comparison. Performance evaluations show \ac{lsdebm} outperforms current leading latent space generative methods, \acp{vae} and \acp{lebm}, in enhancing vertebrae models.


\section{Previous work} \label{sec:prevwork}
Previous approaches have explored enhancing the resulting anatomical model quality from thick slice \ac{mri}, but a large focus has been on \ac{srr} as a preprocessing step in Fig.~\ref{fig:pipeline}~\cite{chai2020mri,zhao2020smore,huang2023super}. \cite{sui2022scan} demonstrates a U-net based 3D approach at the reconstruction step, and high resolution models can be refined from the thick slice reconstructions, for example using \acp{vae} or shape priors in post-processing~\cite{turella2021high,amiranashvili2022learning}. We focus on the last approach of developing post-processing procedures which are integrable into existing pipelines.

We highlight the automated pipeline introduced by \cite{turella2021high} which consists of a segmentation network, the ReconNet, and a \ac{vae}-based post-processing step (Fig.~\ref{fig:pipeline}). Despite their efficiency, \acp{vae} are notorious for generating outputs which are the average of all likely outputs, resulting in something akin to oversmoothing, which is also observed in~\cite{turella2021high}. This can be attributed both to per-pixel loss functions~\cite{houImprovingVariationalAutoencoder2019}, the latent space prior being suboptimal~\cite{odaibo2019tutorial}, and the gap between real and approximate posterior distributions. 

An alternative to \acp{vae} are \acp{ddpm} ~\cite{ho2020denoising}, designed for high-quality image generation. \Acp{ddpm} add noise to training data and then learn the backward denoising process. However, \acp{ddpm} require a full-dimensional, image-scale latent space and a lengthy diffusion process, leading to high computational costs. When it comes to large 3D medical images, this cost can be prohibitively high. 

Other methods that optimize the latent space of \acp{vae} have been shown to improve generated samples and reduce their computational cost. For example, \acp{lebm} replace the encoder of a \ac{vae} by an \ac{ebm} to learn an energy-based latent space via \ac{mcmc} sampling~\cite{pang2020learning,du2019implicit}. Another example is normalizing flows \cite{NEURIPS2023_572a6f16}. Appendix~\ref{app:theory} provides the theoretical basis of our model and covers \ac{ebm} and \ac{ddpm} in detail. Appendix~\ref{app:methods} compares the previous methods.

A similar approach has recently been applied in the field of interpretable text modeling~\cite{yu2022latent}.
Their model focuses on generating creative and varied text outputs, which is encouraged via a symbol-vector coupling which can be used to condition the results. However, this comes at computational cost, which is feasible for their low dimensional data. Our data-driven approach is more suitable for the medical setting and, by avoiding this symbol-vector coupling as explained in Appendix~\ref{app:prevwork}, can be easily applied to high dimensional image data.

\section{Method}\label{sec:method}


\begin{figure}[htb]
\centering
    \includegraphics[width=0.8\textwidth]{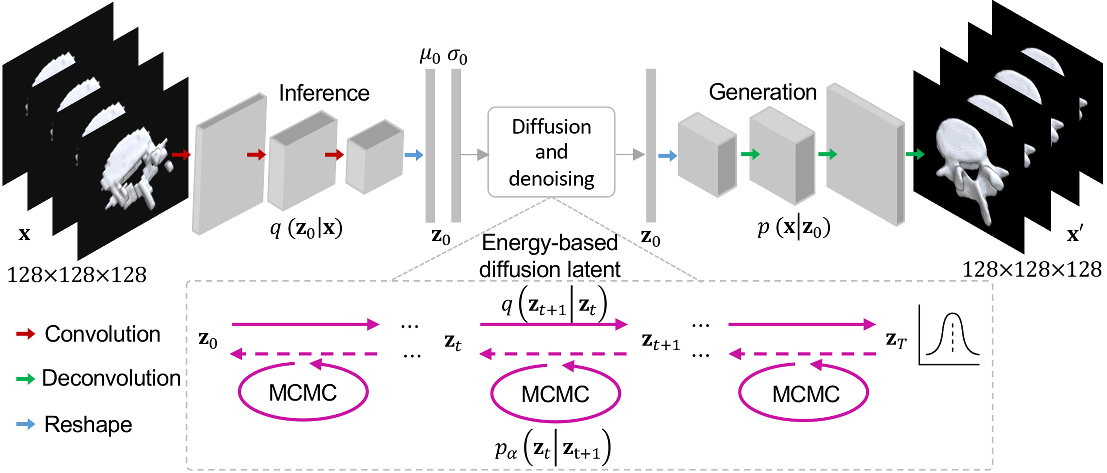}
    \caption{The schematic diagram of our network structure and proposed \ac{lsdebm}. The input is encoded into the latent space $\mathbf{z}$, where a forward diffusion process is constructed and a reverse process with a conditional energy-prior is learned. $\mathbf{z}_0$ is then decoded back into the image dimensions.}
    

\label{fig:method}
\end{figure}

The overall architecture of the \ac{lsdebm} is visualized in Fig.~\ref{fig:method}.
Given an input 3D image $\mathbf{x}$, the inference network generates the latent variable $\mathbf{z}_0 \sim q_\varphi(\mathbf{z}_0|\mathbf{x})= \mathcal{N}(\mathbf{z};\mu_0(x),\sigma_0(x))$ with learnable mean $\mu_0$ and variance $\sigma_0$. A latent diffusion and denoising processes are constructed with the energy-based prior to optimize $\mathbf{z}_0$~\cite{gao2020learning}. The diffusion in latent space acts as checkpoints guiding the learning while also reducing its computational overhead which would be prohibitive in full image space, therefore resulting in more stable and accurate generation. The optimized $\mathbf{z}_0$ is then used by the generation network to reconstruct the 3D image $\mathbf{x}^\prime \sim p_\beta(\mathbf{x}|\mathbf{z}_0)$.
To this end, the latent diffusion process is defined as Markov chain: in Eq.~\eqref{equ:ddpm_markov}~\cite{sohl2015deep}
\begin{equation}
    q(\mathbf{z}_{t+1} | \mathbf{z}_t) := \mathcal{N}(\mathbf{z}_{t+1}; \sqrt{1-\sigma^2_{t+1}}\mathbf{z}_t, \sigma^2_{t+1} \mathbf{I}),
\label{equ:ddpm_forward}
\end{equation}
where $\sigma^2_{t+1}$ is the noise schedule applied to the latent variables in each diffusion step.
The \ac{lsdebm} implements the conditional \ac{ebm}~\cite{gao2020learning}, where a new latent $\tilde{\mathbf{z}}_t = \sqrt{1-\sigma^2_{t+1}}\mathbf{z}_t$ is defined such that its conditional probability, $p_{\alpha} (\tilde{\mathbf{z}}_t | \mathbf{z}_{t+1})$, is described by a Boltzmann distribution:

\begin{equation}
\begin{gathered}
    p_{\alpha} (\tilde{\mathbf{z}}_t | \mathbf{z}_{t+1}) = \frac{\text{exp} \left(-\mathrm{E}_{\alpha}(\tilde{\mathbf{z}}_t, t) - \frac{1}{2\sigma_{t+1}^2}|| \mathbf{z}_{t+1} - \tilde{\mathbf{z}}_t ||^2 \right)}{\tilde{Z}_{\alpha}(\mathbf{z}_{t+1}, t+1)} , \\
    \text{with} \quad \tilde{Z}_{\alpha}(\mathbf{z}_{t+1}, t+1) = \int \text{exp}\left(-\mathrm{E}_{\alpha}(\tilde{\mathbf{z}}_t, t) - \frac{1}{2\sigma_{t+1}^2}|| \mathbf{z}_{t+1} - \tilde{\mathbf{z}}_t ||^2 \right)d\tilde{\mathbf{z}}_t.
\end{gathered}
\label{equ:cond_en_of_x}
\end{equation}
Eq.~\eqref{equ:cond_en_of_x} defines the reverse latent space process where we perform \ac{mcmc} sampling between denoising steps as in Fig.~\ref{fig:method}. Like a vanilla \ac{ebm}, the energy function $\mathrm{E}_{\alpha}$ is parameterized by a neural network. In contrast to the \ac{lebm}, this energy function has an additional time argument due to the quadratic term in the partition function $Z$ which constrains the energy landscape and facilitates sampling~\cite{gao2020learning}. Because $\tilde{\mathbf{z}}_t$ is easily obtained by $\mathbf{z}_t$ from $\mathbb{E}[\mathbf{z}_{t+1}]=\tilde{\mathbf{z}}_t$, in practice $p_{\alpha} (\mathbf{z}_t | \mathbf{z}_{t+1})$ can be used instead of $p_{\alpha}(\tilde{\mathbf{z}}_t | \mathbf{z}_{t+1})$ and is determined using maximum likelihood estimation. We use \ac{mcmc} sampling via Langevin dynamics~\cite{welling2011bayesian}, where

\begin{equation}
    \mathbf{z}^{k+1}_t = \mathbf{z}^k_t - \frac{\lambda}{2}\nabla_\mathbf{z} \log  p_{\alpha} (\mathbf{z}^k_t | \mathbf{z}_{t+1}) + \omega_k, \quad \omega_k \sim \mathcal{N}(0, \lambda), \quad k=1,2,...,K.
\label{equ:mcmc}
\end{equation} 
In practice, $p_{\alpha}$ is approximated by the estimated distribution $q_{\alpha} (\tilde{\mathbf{z}}_t)$. $q_{\alpha} (\tilde{\mathbf{z}}_t) \rightarrow p_{\alpha} (\tilde{\mathbf{z}}_t)$ when the iteration steps $K \rightarrow \infty$ and $\lambda \rightarrow 0$. The gradient of the log likelihood is given by
\begin{equation}
\begin{aligned}
    \nabla_{\mathbf{z}} \log p_{\alpha} (\mathbf{z}_t | \mathbf{z}_{t+1}) = -\nabla_{\mathbf{z}} \mathrm{E}_{\alpha}(\mathbf{z}_t, t) + \frac{1}{\sigma_{t+1}^2} (\mathbf{z}_{t+1} - \mathbf{z}_t),
\end{aligned}
\label{equ:gradient_con_en}
\end{equation}
where $\mathbf{z}_t$ is updated by Eq.~\eqref{equ:mcmc} such that the final latent variable $\mathbf{z}_0$ is obtained at the last step $t=0$. The output image is reconstructed from $\mathbf{z}_0$ using the generation network, i.e., $\mathbf{x}^\prime \sim p_\beta(\mathbf{x}|\mathbf{z}_0) = \mathcal{N}(\beta(\mathbf{z}_0), \sigma I_D)$. Similarly to the \ac{vae}, an Evidence-based Lower BOund (ELBO) can be derived, see Appendix~\ref{app:ELBO}. The encoding and generation networks $\varphi, \beta$ are trained simultaneously for each gradient descent pass to minimize the reconstruction loss of the 3D images. 
We initially validate the generation ability of the method on standard 2D image datasets; additionally, they serve as an initial verification of the generalizability of the method, see Appendix~\ref{app:2D}.

\section{Results}

\subsection{Datasets and Metrics}

We consider two vertebrae reconstruction datasets for 3D vertebrae segmentations from the work of \cite{turella2021high}, including the \ac{ct} based data for the training of the model (denoted as CT-Train, 446 images in total) and paired MRI-CT dataset for the testing of the model, including 80 low-quality MRI images (L-MRI) and corresponding high-quality CT images (H-CT) as ground-truth segmentation. The vertebral masks used for segmentation are of resolution 1mm$^3$, which aligns to the typical lumbar spine protocol resolutions scanned by current MRI scanners. Both datasets consist of $128^3$ binary valued pixel patches from either \ac{ct} or \ac{mri} scans, where the patch size is a result of the cropping of the complete lower lumbar spine into individual vertebrae. 

The evaluation metrics for vertebrae datasets are selected across different categories of measure~\cite{taha2015metrics}, including \ac{dice} for reproducibility, \ac{vs} for similarities of segment volumes, \ac{sen} for the true positive ratio, \ac{spec} for the true negative ratio, \ac{nmi} for the shared information between volumes, and \ac{ck} for inter-annotator agreement between volumes, as defined in Appendix~\ref{app:metrics}~\cite{sklearn_api,muller2022miseval}.

We use this variety of measurements because the available data is very limited, and because the nature of such 3D medical data makes it challenging for a single metric to accurately capture similarity. For example, the \ac{vae} used in the post-processing of~\cite{turella2021high} gives a good \ac{dice} score, but nevertheless the reconstructions are unrealistically smooth. Additionally, the variance of the latent space is used to understand the latent space priors learned by the \ac{lsdebm} and \ac{lebm}. Implementation details can be found in Appendix~\ref{app:implementation}.

\subsection{Lumbar Vertebrae Reconstruction}

\begin{figure}[!ht]
\centering
\begin{minipage}{0.24\textwidth}
    \centering
    \includegraphics[width=\linewidth]{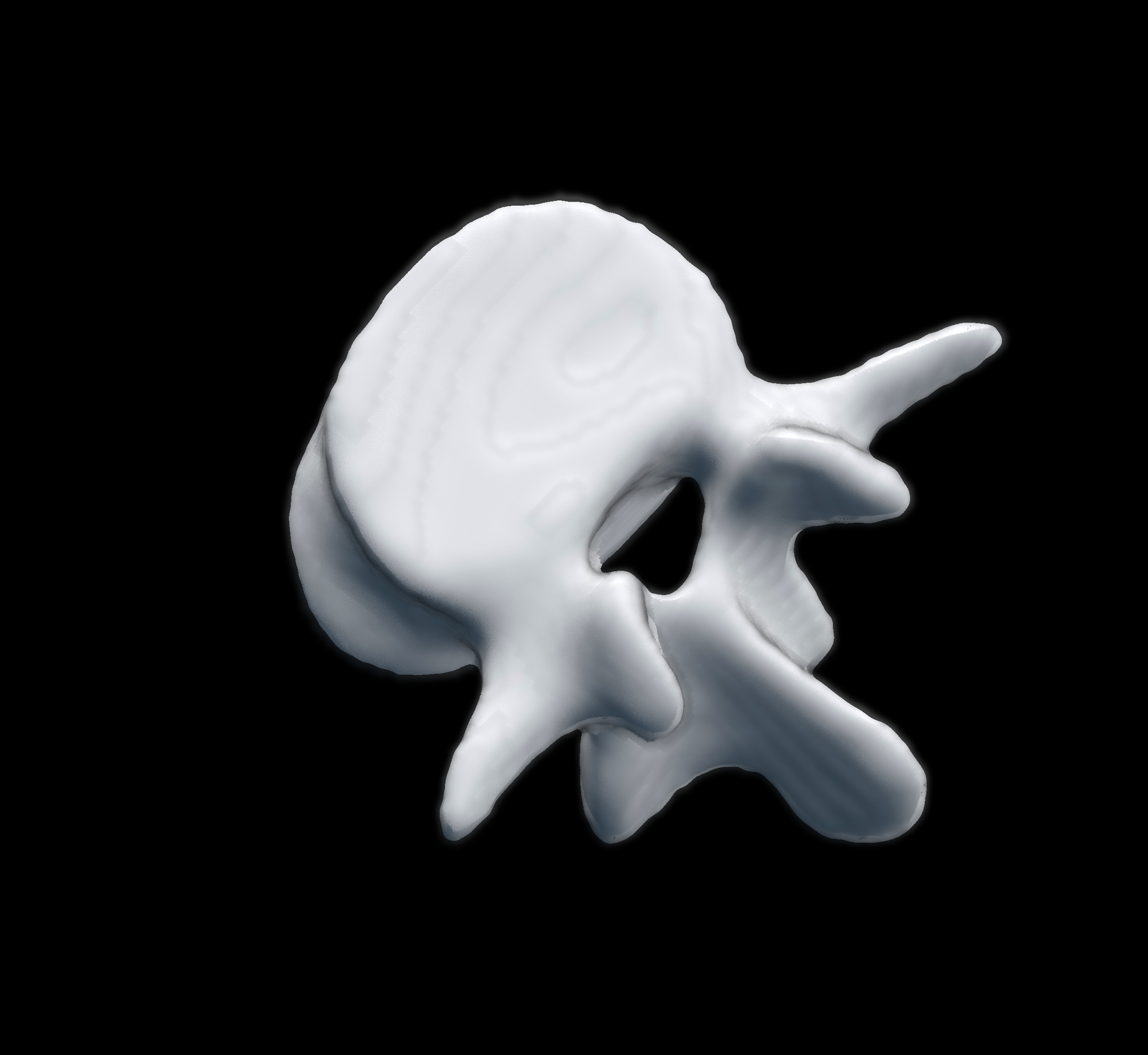}
    \caption*{\ac{vae}}
\end{minipage}
\begin{minipage}{0.24\textwidth}
    \centering
    \includegraphics[width=\linewidth]{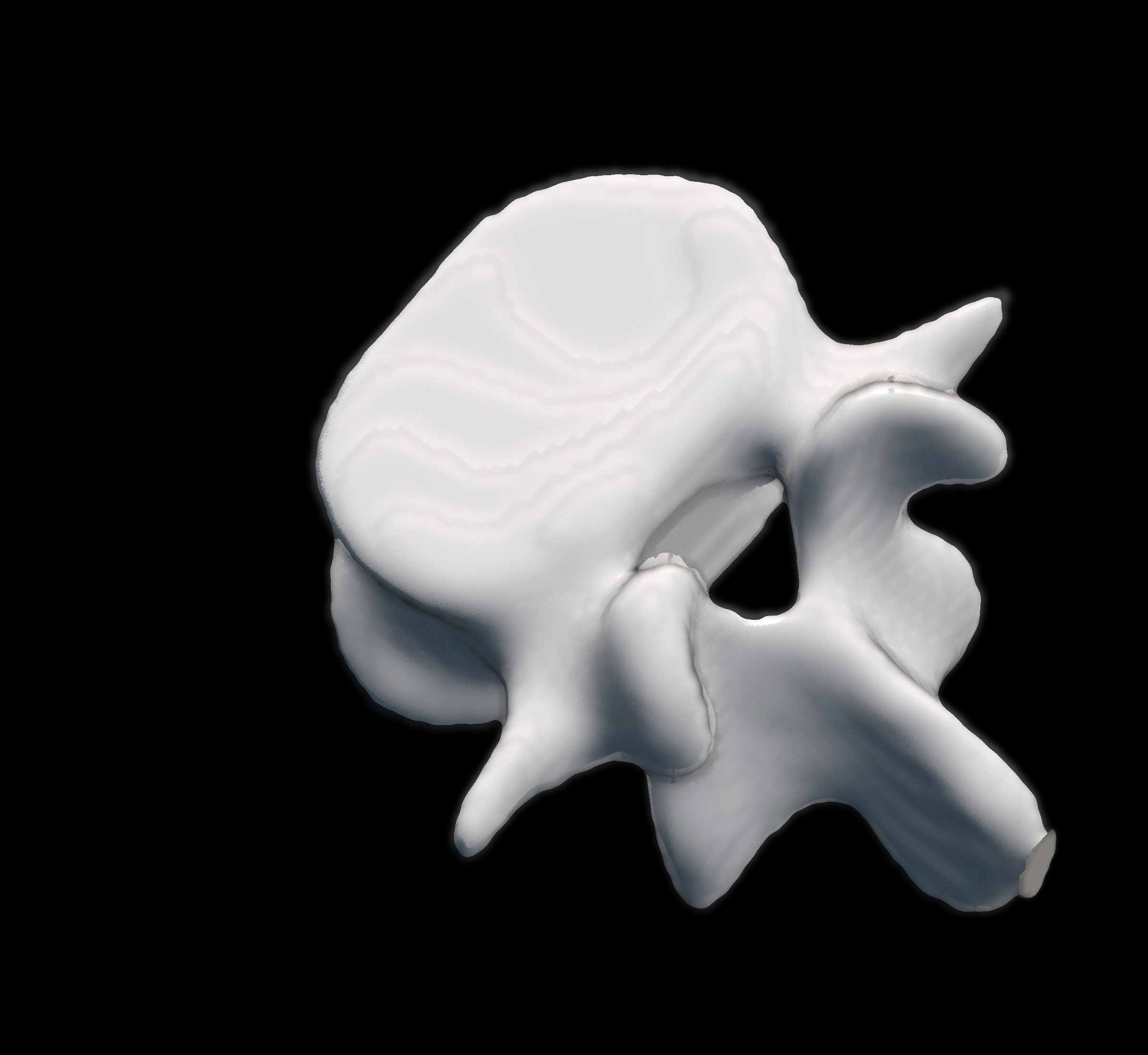}
    \caption*{\ac{lebm}}
\end{minipage}
\begin{minipage}{0.24\textwidth}
    \centering
    \includegraphics[width=\linewidth]{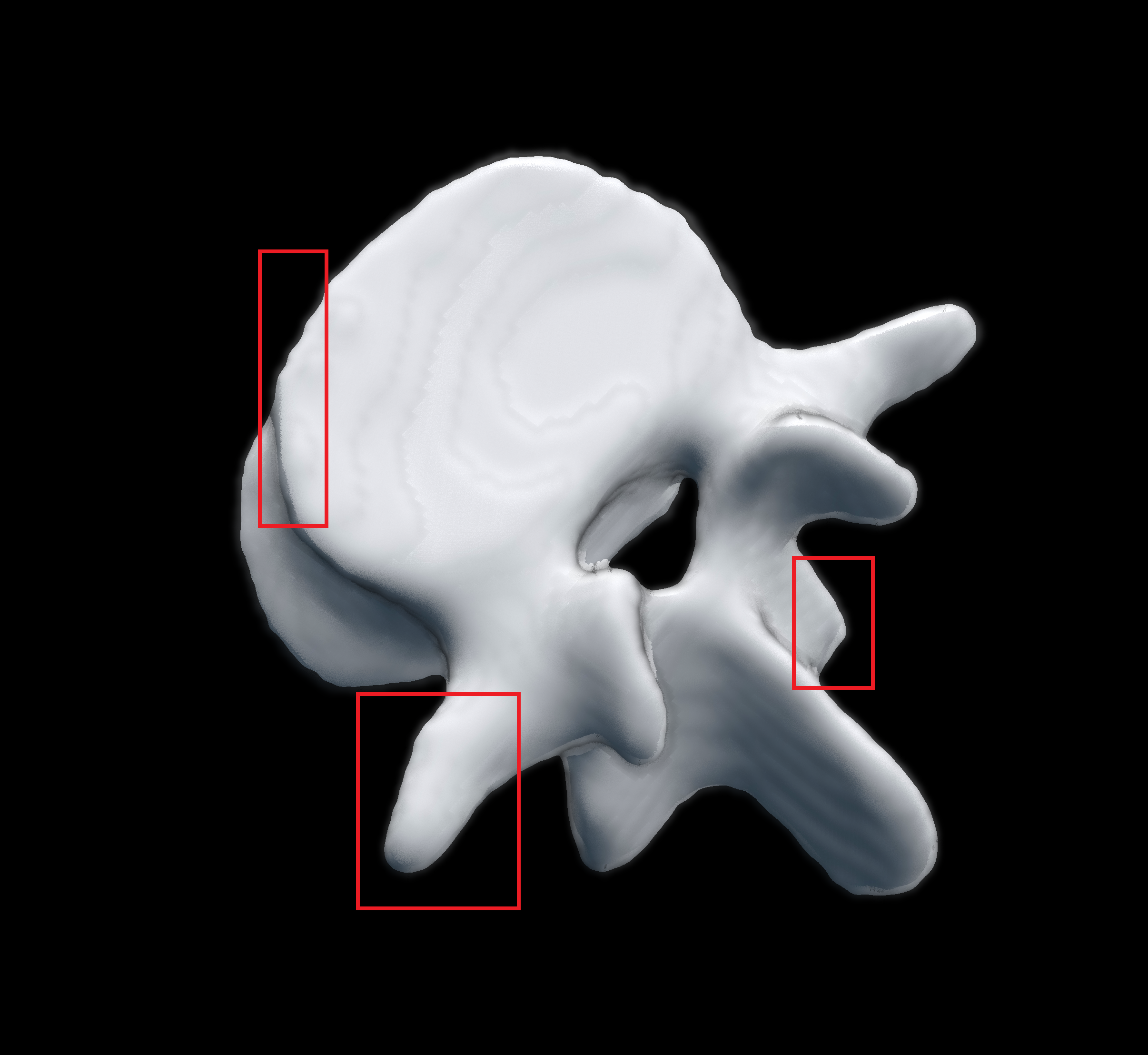}
    \caption*{\ac{lsdebm}}
\end{minipage}
\begin{minipage}{0.24\textwidth}
    \centering
    \includegraphics[width=\linewidth]{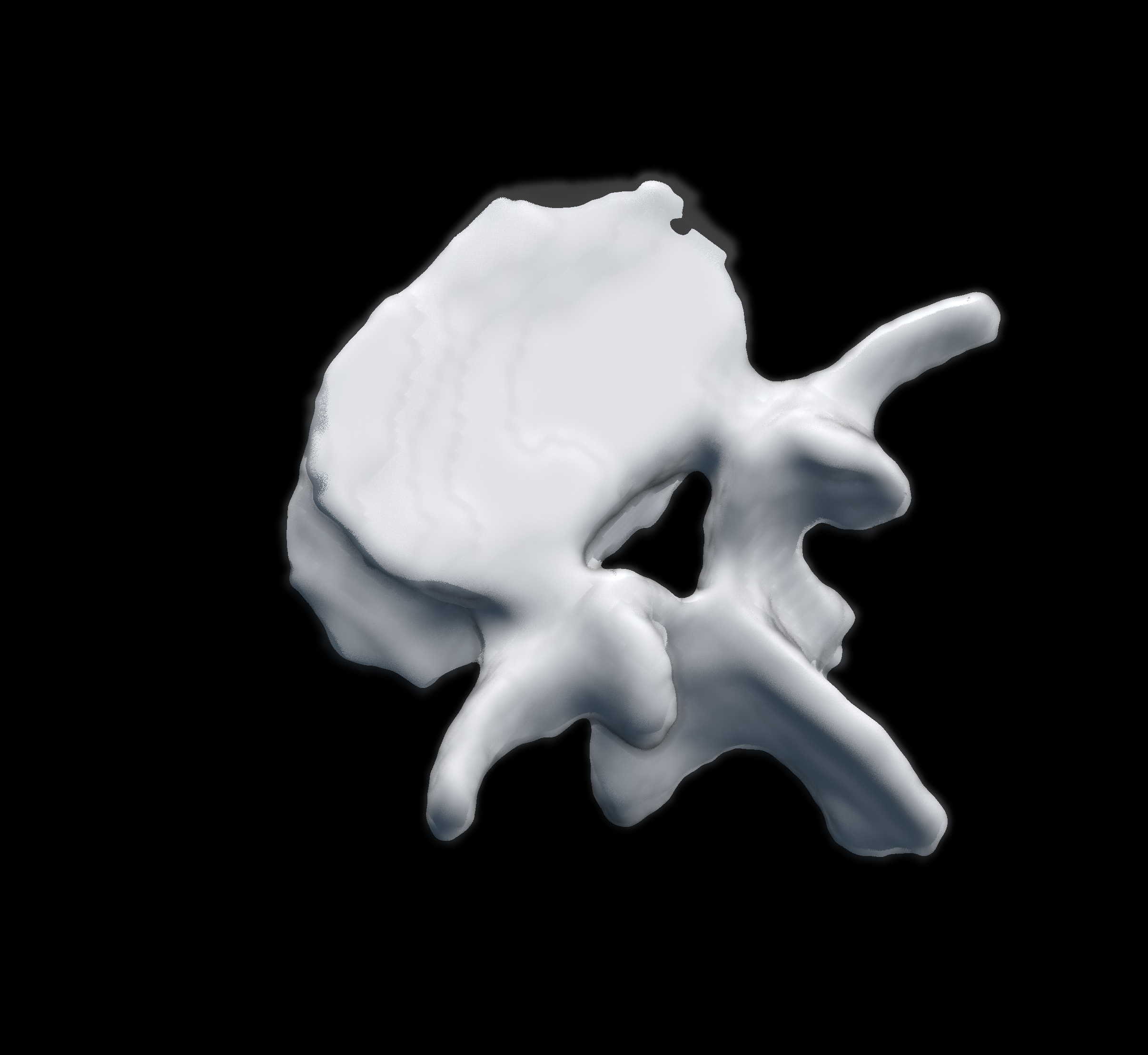}
    \caption*{H-CT}
\end{minipage}
\caption{Comparison of \ac{vae}, \ac{lebm}, and \ac{lsdebm} reconstructions of the L3 vertebra, where H-CT represents the high-quality CT image ground truth. The red boxes denote regions of interest for qualitative comparison. The \ac{lsdebm}'s reconstruction is more faithful to H-CT.}
\label{fig:models_l3}
\end{figure}

\begin{table}[htb]
\caption{Comparison of \ac{vae}, \ac{lebm}, and \ac{lsdebm} on the L-MRI dataset. The mean $\pm$ standard deviation are taken across 80 test set samples.}
\setlength{\tabcolsep}{3pt}
\label{tab:vert_lq}
\begin{adjustbox}{center}
\begin{tabular}{|p{45pt}|p{45pt}|p{45pt}|p{45pt}|p{45pt}|p{45pt}|p{45pt}|}
\hline
Method & \ac{dice} & \ac{vs} & \ac{sen} & \ac{spec} & \ac{nmi} & \ac{ck}\\
\hline
\ac{vae} & \makecell[l]{0.7626\\($\pm$0.0457)}
        & \makecell[l]{0.7887\\($\pm$0.0448)} 
        & \makecell[l]{0.9667\\($\pm$0.0138)}
        & \makecell[l]{0.9882\\($\pm$0.0026)}
        & \makecell[l]{0.6252\\($\pm$0.0451)}
        & \makecell[l]{0.7566\\($\pm$0.0461)}\\
\ac{lebm} & \makecell[l]{0.7619\\($\pm$0.0576)}
        & \makecell[l]{0.7866\\($\pm$0.0539)} 
        & \makecell[l]{\textbf{0.9692}\\($\pm$0.0610)}
        & \makecell[l]{0.9883\\($\pm$0.0026)}
        & \makecell[l]{0.6304\\($\pm$0.0663)}
        & \makecell[l]{0.7560\\($\pm$0.0583)}\\     
\ac{lsdebm} & \makecell[l]{\textbf{0.8304}\\($\pm$ 0.0317)}
        & \makecell[l]{\textbf{0.8627}\\($\pm$ 0.0313)} 
        & \makecell[l]{0.9625 \\($\pm$ 0.0135)}
        & \makecell[l]{\textbf{0.9914}\\($\pm$0.0020)}
        & \makecell[l]{\textbf{0.6973}\\($\pm$0.0367)}
        & \makecell[l]{\textbf{0.8258}\\($\pm$0.0321)}\\ 
\hline
\end{tabular}
\end{adjustbox}
\end{table}

The methods \ac{vae}, \ac{lebm}, and \ac{lsdebm} are trained using \ac{ct}-Train images and then applied to low-quality \ac{mri} images (L-\ac{mri}) to generate missing details. We evaluate these methods by comparing the reconstructed \ac{mri} vertebrae from L-MRI with those reconstructed from H-CT, with metrics in Table~\ref{tab:vert_lq} and sample reconstructions in Fig.~\ref{fig:models_l3}. Additional reconstructions are shown in Appendix~\ref{app:3D}.

Our \ac{lsdebm} outperforms the \ac{vae} and \ac{lebm} in terms of \ac{dice} and \ac{vs} scores, indicating better reproducibility and higher similarity to the H-CT volume. Although \ac{lebm} achieves a higher \ac{sen} score, it scores lower in \ac{spec}, which suggests that \ac{lebm} generates fewer false negative segments, but may miss some image details. This observation is also supported qualitatively by the red boxes in Fig.~\ref{fig:models_l3}, highlighting missing detail in both the \ac{vae} and \ac{lebm} outputs. Additionally, \ac{lsdebm}'s superior \ac{nmi} and \ac{ck} scores indicate that \ac{lsdebm}'s reconstructions share more information with the H-CT volume. 

\begin{figure}[htb]
\centering
\includegraphics[width=0.8\textwidth]{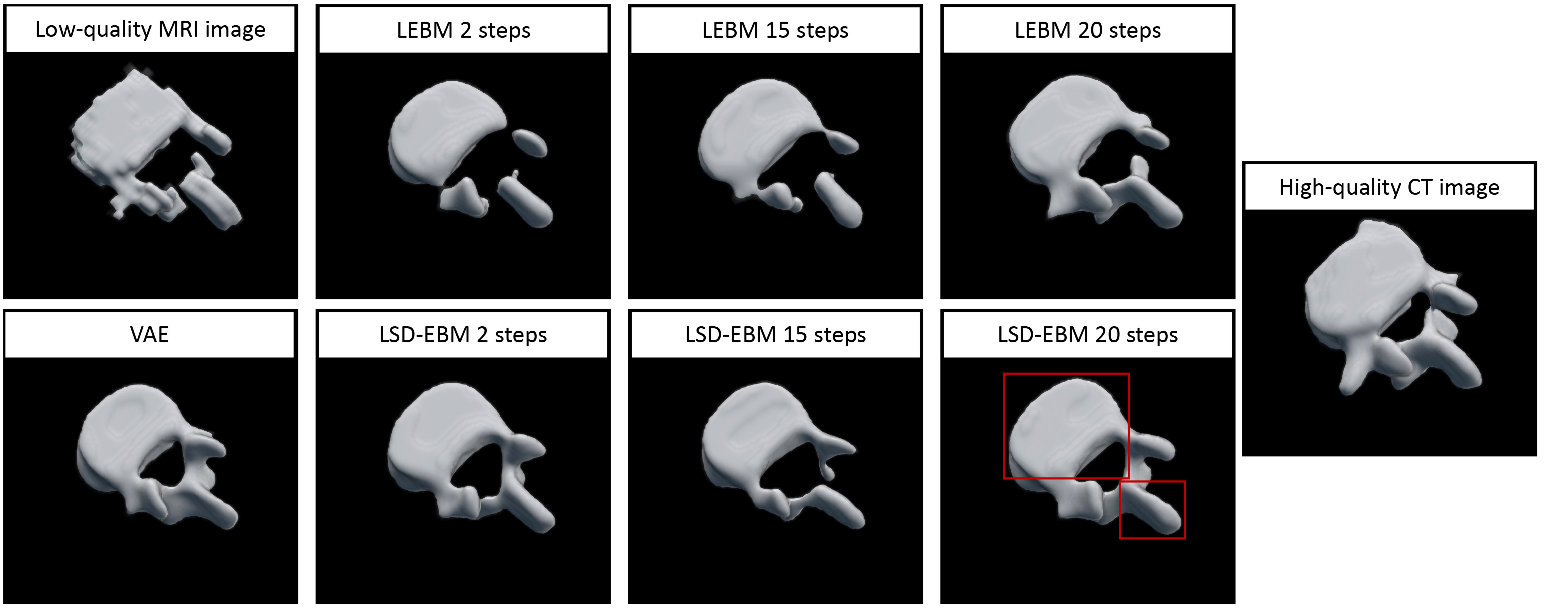}
\caption{The visualization of \ac{vae}, \ac{lebm}, and \ac{lsdebm} on the reconstruction results of low-quality \ac{mri} with reference to the high-quality \ac{ct} image on the right.  For the \ac{lebm}, and \ac{lsdebm} the intermediate reconstructions from the latent space at  2, 15, and 20 time steps are also shown. The red boxes denote regions of interest for qualitative comparison.} 
\label{fig:lq2hq}
\end{figure}

As seen in Fig.~\ref{fig:lq2hq}, \ac{vae} tends to smooth out finer features more than \ac{lsdebm} and H-CT images. \ac{lebm}, with 20 sampling steps, reveals more details than \ac{vae} but still underperforms if using fewer steps. In contrast, \ac{lsdebm} consistently retains detailed features across various sampling steps. This demonstrates that \ac{lsdebm}'s sampling process is more stable than \ac{lebm}'s due to the denoising optimization process for the energy-based prior. This also allows for efficient model utilization with fewer steps. The \ac{lsdebm}'s high performance comes with lower time complexity, training in 17h as compared to 12h for the \ac{vae} and 33h for the \ac{lebm} (Appendix~\ref{app:time}). The processing of \ac{ddpm} with just two steps exceeded the 40 GB GPU memory limit, highlighting its computational inefficiency. The \ac{lsdebm} is shown to efficiently reconstruct more faithful, higher-quality \ac{mri} vertebrae segmentations compared to the \ac{vae} and \ac{lebm}.

\subsection{Convergence in the Latent Space}

\begin{figure}[htb]
 \centering
 \includegraphics[width=1\textwidth]{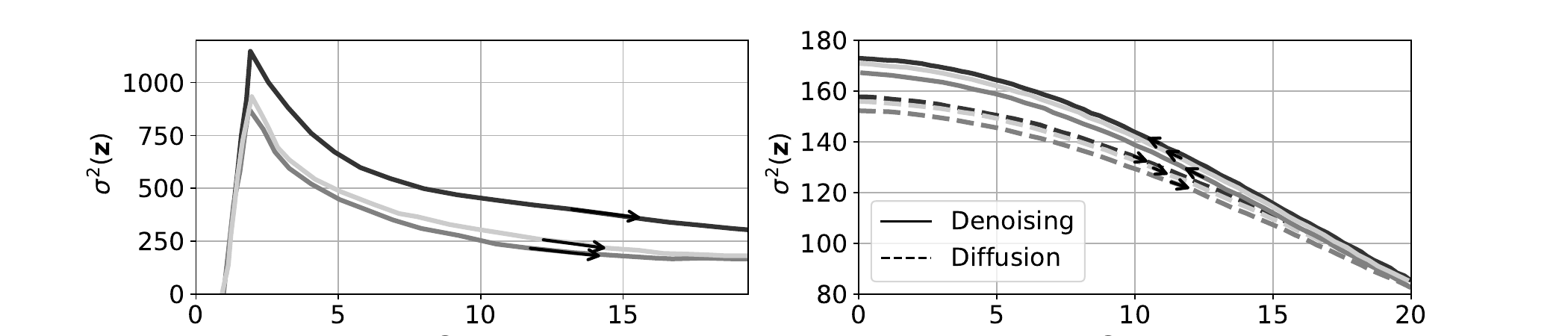}
 \caption{The mean variance of the latent variables (\textbf{left}) of the \ac{mcmc} sampling process in \ac{lebm} and (\textbf{right}) the diffusion and denoising processes in \ac{lsdebm}. The different shades represent repetitions. The arrows denote the time direction of the respective process.}
 \label{fig:latent}
\end{figure}

We analyze the variance of the latent variables at each step in Fig.~\ref{fig:latent}, which serves as a measure for how much a sample resembles random noise. A lower variance is attributed to less noise and therefore a stronger learned signal.

In the \ac{lebm} case, the shape of the variance exhibits an expected behaviour: the spike followed by a gradual convergence to a variable minimum matches the burn-in or calibration followed by convergence period of \ac{mcmc} methods. The \ac{lsdebm}, in contrast, converges directly and with more consistency across runs to a comparatively lower variance. Its stability is a direct result of the well-defined denoising process, detailed in~\cite{ho2020denoising} and described in Sec.~\ref{sec:method}. This ensures the consistency of learned latent spaces across different runs and facilitates better reconstructions at various time-steps, as evidenced in Fig.~\ref{fig:lq2hq}.

\section{Conclusions}

In this study, we enhanced the quality of low-quality \ac{mri} vertebra models from thick-slice images. We develop and implement a latent energy-based model trained on high-quality \ac{ct} data, \ac{lsdebm}, which demonstrated superior reconstructions compared to \acp{vae} and \acp{lebm}. It not only addressed the computational challenges in diffusion models, making them suitable for the 3D medical imaging regime, but also enhanced reconstruction performance. Furthermore, our model exhibited a more stable generative process with a comparable time cost to \acp{vae}, taking half as long as the \ac{lebm}. Although our method relies on high resolution domain specific \ac{ct} images, our results bolster the feasibility of using \ac{mri} as an efficient and safer alternative to \ac{ct} scans in vertebrae modeling. Future work will include understanding latent feature extraction for domain adaptation and generalizability.



\subsubsection{Acknowledgments} This project was supported by grant \#2022-643 of the Strategic Focus Area "Personalized Health and Related Technologies (PHRT)" of the ETH Domain (Swiss Federal Institutes of Technology).

\begin{credits}

\subsubsection{\discintname}
The authors have no competing interests to declare that are
relevant to the content of this article.
\end{credits}
%
%
%
 \bibliographystyle{splncs04}
\bibliography{reference}

\begin{thebibliography}{10}
\providecommand{\url}[1]{\texttt{#1}}
\providecommand{\urlprefix}{URL }
\providecommand{\doi}[1]{https://doi.org/#1}

\bibitem{ackley1985}
Ackley, D.H., Hinton, G.E., Sejnowski, T.J.: A learning algorithm for boltzmann
  machines. Cognitive Science  \textbf{9}(1),  147--169 (1985).
  \doi{https://doi.org/10.1016/S0364-0213(85)80012-4},
  \url{https://www.sciencedirect.com/science/article/pii/S0364021385800124}

\bibitem{amiranashvili2022learning}
Amiranashvili, T., L{\"u}dke, D., Li, H.B., Menze, B., Zachow, S.: Learning
  shape reconstruction from sparse measurements with neural implicit functions.
  In: International Conference on Medical Imaging with Deep Learning. pp.
  22--34. {PMLR} (2022)

\bibitem{bajger2021lumbar}
Bajger, M., To, M.S., Lee, G., Wells, A., Chong, C., Agzarian, M., Poonnoose,
  S.: Lumbar spine {{CT}} synthesis from {{MR}} images using {{CycleGAN-a}}
  preliminary study. In: Digital Image Computing: {{Techniques}} and
  Applications ({{DICTA}}). pp.~1--8. {IEEE} (2021)

\bibitem{been2010new}
Been, E., Barash, A., Pessah, H., Peleg, S.: A new look at the geometry of the
  lumbar spine. Spine (Philadelphia, Pa. : 1986)  \textbf{35}(20),
  E1014--E1017 (2010)

\bibitem{sklearn_api}
Buitinck, L., Louppe, G., Blondel, M., Pedregosa, F., Mueller, A., Grisel, O.,
  Niculae, V., Prettenhofer, P., Gramfort, A., Grobler, J., Layton, R.,
  VanderPlas, J., Joly, A., Holt, B., Varoquaux, G.: {{API}} design for machine
  learning software: Experiences from the scikit-learn project. In: {{ECML
  PKDD}} Workshop: {{Languages}} for Data Mining and Machine Learning. pp.
  108--122 (2013)

\bibitem{chai2020mri}
Chai, Y., Xu, B., Zhang, K., Lepore, N., Wood, J.C.: {{MRI}} restoration using
  edge-guided adversarial learning. IEEE Access : practical innovations, open
  solutions  \textbf{8},  83858--83870 (2020)

\bibitem{du2019implicit}
Du, Y., Mordatch, I.: Implicit generation and modeling with energy based
  models. In: Wallach, H., Larochelle, H., Beygelzimer, A., {dAlch{\'e}-Buc},
  F., Fox, E., Garnett, R. (eds.) Advances in Neural Information Processing
  Systems. vol.~32. {Curran Associates, Inc.} (2019)

\bibitem{NEURIPS2023_572a6f16}
Flouris, K., Konukoglu, E.: Canonical normalizing flows for manifold learning.
  In: Oh, A., Neumann, T., Globerson, A., Saenko, K., Hardt, M., Levine, S.
  (eds.) Advances in Neural Information Processing Systems. vol.~36, pp.
  27294--27314. Curran Associates, Inc. (2023),
  \url{https://proceedings.neurips.cc/paper_files/paper/2023/file/572a6f16ec44f794fb3e0f8a310acbc6-Paper-Conference.pdf}

\bibitem{gao2020learning}
Gao, R., Song, Y., Poole, B., Wu, Y.N., Kingma, D.P.: Learning energy-based
  models by diffusion recovery likelihood. arXiv preprint arXiv:2012.08125
  (2020)

\bibitem{ho2020denoising}
Ho, J., Jain, A., Abbeel, P.: Denoising diffusion probabilistic models.
  Advances in Neural Information Processing Systems  \textbf{33},  6840--6851
  (2020)

\bibitem{Hopfield1982NeuralNA}
Hopfield, J.J.: Neural networks and physical systems with emergent collective
  computational abilities. Proceedings of the National Academy of Sciences of
  the United States of America  \textbf{79 8},  2554--8 (1982),
  \url{https://api.semanticscholar.org/CorpusID:784288}

\bibitem{houImprovingVariationalAutoencoder2019}
Hou, X., Sun, K., Shen, L., Qiu, G.: Improving variational autoencoder with
  deep feature consistent and generative adversarial training. Neurocomputing
  \textbf{341},  183--194 (May 2019). \doi{10.1016/j.neucom.2019.03.013}

\bibitem{huang2023super}
Huang, S., Chen, G., Sun, K., Cui, Z., Zhang, X., Xue, P., Zhang, X., Zhang,
  H., Shen, D.: Super-resolution reconstruction of fetal brain {{MRI}} with
  prior anatomical knowledge. In: International Conference on Information
  Processing in Medical Imaging. pp. 428--441. {Springer} (2023)

\bibitem{laakso1997mri}
Laakso, M.P., Juottonen, K., Partanen, K., Vainio, P., Soininen, H.: {{MRI}}
  volumetry of the hippocampus: The effect of slice thickness on volume
  formation. Magnetic Resonance Imaging  \textbf{15}(2),  263--265 (1997)

\bibitem{muller2022miseval}
M{\"u}ller, D., Hartmann, D., Meyer, P., Auer, F., {Soto-Rey}, I., Kramer, F.:
  {{MISeval}}: A metric library for medical image segmentation evaluation.
  Challenges of trustable AI and added-value on health. Proceedings of MIE
  (2022)

\bibitem{odaibo2019tutorial}
Odaibo, S.: Tutorial: {{Deriving}} the standard variational autoencoder (vae)
  loss function. arXiv preprint arXiv:1907.08956  (2019)

\bibitem{pang2020learning}
Pang, B., Han, T., Nijkamp, E., Zhu, S.C., Wu, Y.N.: Learning latent space
  energy-based prior model. Advances in Neural Information Processing Systems
  \textbf{33},  21994--22008 (2020)

\bibitem{sohl2015deep}
{Sohl-Dickstein}, J., Weiss, E., Maheswaranathan, N., Ganguli, S.: Deep
  unsupervised learning using nonequilibrium thermodynamics. In: International
  Conference on Machine Learning. pp. 2256--2265. {PMLR} (2015)

\bibitem{sui2022scan}
Sui, Y., Afacan, O., Jaimes, C., Gholipour, A., Warfield, S.K.:
  Scan-{{Specific}} generative neural network for {{MRI}} super-resolution
  reconstruction. IEEE Transactions on Medical Imaging  \textbf{41}(6),
  1383--1399 (2022)

\bibitem{taha2015metrics}
Taha, A.A., Hanbury, A.: Metrics for evaluating {{3D}} medical image
  segmentation: Analysis, selection, and tool. BMC Medical Imaging
  \textbf{15}(1),  1--28 (2015)

\bibitem{turella2021high}
Turella, F., Bredell, G., Okupnik, A., Caprara, S., Graf, D., Sutter, R.,
  Konukoglu, E.: High-resolution segmentation of lumbar vertebrae from
  conventional thick slice mri. In: Medical Image Computing and Computer
  Assisted {{Intervention}}{\textendash}{{MICCAI}} 2021: 24th International
  Conference, Strasbourg, France, September 27{\textendash}{{October}} 1, 2021,
  Proceedings, Part {{I}} 24. pp. 689--698. {Springer} (2021)

\bibitem{turner2005cd}
Turner, R.:  (2005)

\bibitem{welling2011bayesian}
Welling, M., Teh, Y.W.: Bayesian learning via stochastic gradient {{Langevin}}
  dynamics. In: Proceedings of the 28th International Conference on Machine
  Learning ({{ICML-11}}). pp. 681--688 (2011)

\bibitem{wuGlobalLowBack2020}
Wu, A., March, L., Zheng, X., Huang, J., Wang, X., Zhao, J., Blyth, F.M.,
  Smith, E., Buchbinder, R., Hoy, D.: Global low back pain prevalence and years
  lived with disability from 1990 to 2017: Estimates from the {{Global Burden}}
  of {{Disease Study}} 2017. Annals of Translational Medicine  \textbf{8}(6),
  ~299 (Mar 2020). \doi{10.21037/atm.2020.02.175}

\bibitem{yu2022latent}
Yu, P., Xie, S., Ma, X., Jia, B., Pang, B., Gao, R., Zhu, Y., Zhu, S.C., Wu,
  {\relax YN}.: Latent diffusion energy-based model for interpretable text
  modeling. In: International Conference on Machine Learning ({{ICML}}). (2022)

\bibitem{zhao2020smore}
Zhao, C., Dewey, B.E., Pham, D.L., Calabresi, P.A., Reich, D.S., Prince, J.L.:
  {{SMORE}}: A self-supervised anti-aliasing and super-resolution algorithm for
  {{MRI}} using deep learning. IEEE Transactions on Medical Imaging
  \textbf{40}(3),  805--817 (2020)

\end{thebibliography}

\newpage
\appendix

\begin{center}
\Large
\textbf{Appendix}
\end{center}

\section{Theoretical Background} \label{app:theory}

\subsection{Energy-based Models}
\label{sec:pre_en}

Energy-based models have a long history tracing back to statistical physics, Hopfield networks~\cite{Hopfield1982NeuralNA}  and Boltzmann machines~\cite{ackley1985}. 

The main principle is to model the prior as an energy function, which can assign an energy to each input sample $\mathbf{x}$ from the data space $\mathcal{X}$. \cite{du2019implicit} uses the \ac{ebm} for image generation, maximizing the likelihood between generated and real image instances by assigning low energy values for realistic images (positive samples) and increasing the energy for unrealistic images (negative samples). The prior can be used in a maximum likelihood estimation method for generation, for example using \ac{mcmc} sampling.

For $\mathcal{X}$ being the distribution for each datum $\mathbf{x} \sim p_{\mathcal{X}}(\mathbf{x})$, the energy function $\mathrm{E}_{\theta} (\mathbf{x})$ can be parameterized by $\theta$, where $\theta$ can be neural network parameters. The energy function defines a probability distribution via the Boltzmann distribution, i.e. models a density over the input space
\begin{equation}
    p_{\theta} (\mathbf{x}) = \frac{\exp\big(-\mathrm{E}_{\theta}(\mathbf{x})\big)}{Z(\theta)}, \quad \text{with} \enspace Z(\theta) = \int \exp(-\mathrm{E}_{\theta}(\mathbf{x}))d\mathbf{x},
\label{equ:en_of_x}
\end{equation}
where $Z(\theta)$ is the normalizing factor.

The objective of the \ac{ebm} tries to maximize the negative log likelihood of $p_{\mathcal{X}}(\mathbf{x})$ as follows
\begin{equation}
    \mathcal{L}(\theta) = \mathbb{E}_{\mathbf{x} \sim p_{\mathcal{X}}(\mathbf{x})}[-\text{log}p_{\theta} (\mathbf{x})] = \mathbb{E}_{\mathbf{x} \sim p_{\mathcal{X}}(\mathbf{x})}[\mathrm{E}_{\theta}(\mathbf{x})-\text{log}Z(\theta)].
\label{equ:nlog_p_e}
\end{equation}
The normalizing factor is of course intractable but the optimization can be performed via a gradient decent. The gradient can be shown to obtain the following form~\cite{turner2005cd} as 
\begin{equation}
    \nabla \mathcal{L}(\theta) \approx  \mathbb{E}_{\mathbf{x}^- \sim p_{\theta}} [-\nabla_{\theta}\mathrm{E}_{\theta} (\mathbf{x}^-)] -\mathbb{E}_{\mathbf{x}^+ \sim p_{\mathcal{X}}} [-\nabla_{\theta}\mathrm{E}_{\theta} (\mathbf{x}^+)],
\label{equ:delta_log_p_e}
\end{equation}
 This gradient decreases the energy of the positive data samples $\mathbf{x}^+ \sim p_{\mathcal{X}} \approx q_{\theta}$\footnote{The data distribution needs to be approximated by a parametric function.}. 
, while increasing the energy of the negative samples $\mathbf{x}^-\sim p_{\theta}$. The sampling can be performed via Langevin dynamics making use of the gradient of the energy function:
\begin{equation}
\tilde{\mathbf{x}}_{k+1} = \tilde{\mathbf{x}}_k - \frac{\lambda}{2}\nabla_\mathbf{x}\mathrm{E}_{\theta} (\tilde{\mathbf{x}}_k) + \omega_k, \quad \omega_k \sim \mathcal{N}(0, \lambda), \quad k=1,2,...,K.
\label{equ:langevin_ebm}
\end{equation}
The iterative procedure defines the estimated distribution $q_{\theta}$ given that $\tilde{\mathbf{x}}_k \sim q_{\theta}$ and as $K \rightarrow \infty$ and $\lambda \rightarrow 0$, $q_{\theta} (\tilde{\mathbf{x}}) \rightarrow p_{\theta} (\mathbf{x})$.

Although, \acp{ebm} have been showcased to faithfully generate images, solving Langevin type equations for a full dimensional image space still remains computationally cumbersome and arguably less expressive than a lower dimensional latent space method. To that effect, \cite{pang2020learning} proposes the \ac{lebm}, a model based on \acp{vae} where the encoding procedure is replaced by a latent space model. Defining the prior as $p_{\alpha} (z)$ with parameters $\alpha$ and a decoder as $p_{\beta} (x|z)$ with parameters $\beta$, the joint probability distribution is formulated as

\begin{equation}
    p_{\theta} (x, z) = p_{\beta} (x|z)p_{\alpha} (z),
\end{equation}

where $p_{\alpha} (z)$ is the energy-based prior of the latent space $z$ and is defined similar to~\eqref{equ:en_of_x},

\begin{equation}
p_{\alpha} (z) = \frac{\exp \big(\mathrm{E}_{\alpha}(z)\big)}{Z(\alpha)}p_0(z), \quad \text{with} \enspace Z(\alpha) = \int \exp(\mathrm{E}_{\alpha}(z))p_0(z)dz.
\label{equ:en_prior}
\end{equation}

where $p_0(z)$ is a base prior, for example, a multivariate normal distribution. Langevin dynamics are then performed for the latent variables, similarly to~\eqref{equ:langevin_ebm}, given the likelihood derivatives can be obtained for the prior and decoder models as explained in~\cite{pang2020learning}.

\subsection{Diffusion Probabilistic Models}
\label{sec:pre_diff}

\cite{sohl2015deep} introduces the \ac{dpm}, which artificially decreases the quality of the data by adding
increasing levels of noise, while training a model to reverse this process, both can be modeled with a Markov chain.
The trained model can be used to generate new samples starting from pure noise. \cite{ho2020denoising} proposes \ac{ddpm} which achieves remarkable results in image synthesis by fixing the variance and learning noise directly. 

The forward - also diffusion or noising - process starts with a data sample from a real distribution \(\mathbf{x} \sim q(\mathbf{x}_0)\), and Gaussian noise is added gradually to the sample in T steps, effectively creating  a Markov chain $\mathbf{x}_1, ..., \mathbf{x}_T$.

\begin{equation}\label{equ:ddpm_markov}
    q(\mathbf{x}_{t+1} | \mathbf{x}_{t}) := \mathcal{N}(\mathbf{x}_{t+1}; \sqrt{1-\sigma^2_{t+1}}\mathbf{x}_t, \sigma^2_{t+1} \mathbf{I}),
\end{equation}
where $\sigma^2_{t+1}$ is the variance schedule of the predefined Gaussian noise. 

The reverse - also the denoising or generative - process, aims to invert the forward diffusion process. Generated samples from the original data distribution are obtained by initiating the forward process with a random noise $\mathbf{x}_T \sim \mathcal{N} ( 0, I )$. Subsequently, running the reverse process reconstructs samples that closely resemble the original data distribution. Parameterizing a model, $\theta$ to approximate the data distribution, we obtain the following:
\begin{equation}
    p_{\theta}(\mathbf{x}_t | \mathbf{x}_{t+1}) := \mathcal{N}(\mathbf{x}_t; \mu_{\theta}(\mathbf{x}_{t+1}, t+1), \Sigma_{\theta}(\mathbf{x}_{t+1}, t+1))
\end{equation}
where $\mu_{\theta}$ and $\Sigma_{\theta}$ can be modeled with a neural network. The objective of a \ac{ddpm} is to maximize the likelihood between the diffusion process step \(q(\mathbf{x}_t |\mathbf{x}_{t+1}, \mathbf{x}_0)\) and denoising process step \(p_{\theta}(\mathbf{x}_t |\mathbf{x}_{t+1})\).


In practice, a neural network is used to predict $\mu_{\theta}$ with fixed term $\Sigma_{\theta}$, reducing the complexity and improving training efficiency. 

\subsection{Comparison of Existing Methods \label{app:methods}
}
\begin{figure}[!ht]
\centering
    \includegraphics{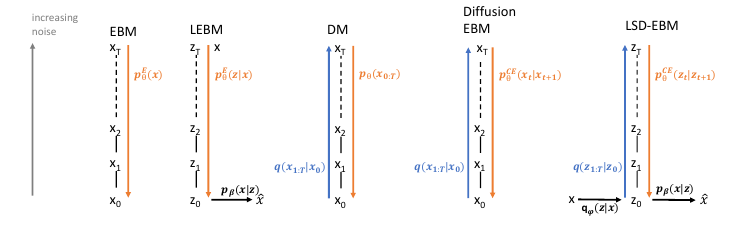}
    \caption{Schematics of different existing methods, with focus on the processes increasing or decreasing the noise in the data and its quality. In order left to right, \ac{ebm}, \ac{lebm}, diffusion model, diffusion \acp{ebm}  and ours. The blue and orange arrows indicate the forward and backward processes respectively, in constant dimension. The black arrows indicate an encoder or decoder depending on their location, and Greek letters indicate a parameter space. $p^{E}$ and $p^{CE}$ mark processes based on \acp{ebm} and conditional \acp{ebm} respectively.}
\label{fig:ebmmethods}
\end{figure}

Some existing methods are compared in Fig.~\ref{fig:ebmmethods}. The \ac{vae} is a variational inference method with an encoder and a decoder; the encoder can have multiple implementations but it is nowadays standard practice to implement it as a neural network. The \ac{lebm} is based on the \ac{vae} and replaces the encoder with an \ac{mcmc} sampling of an energy based prior. \acp{vae} often assume Gaussian priors and posteriors, while \acp{lebm} offer more flexibility in defining the energy function. Our method implements an autoencoder-like architecture which brings in the performance capability of diffusion models and combines them with \acp{ebm}. However, in contrast to \acp{lebm}, we implement the conditional \ac{ebm} in the latent space which is less computationally expensive, and find that this choice and our diffusion-like architecture results in more accurate and efficient reconstructions. 

\section{Further Previous Work}\label{app:prevwork}
\cite{yu2022latent} relies on an information bottleneck in conjunction with geometric clustering for their symbol-vector coupling to avoid mode-collapse and generates more creative text outputs. Their symbol-vector coupling \ac{ebm} results in the distribution $p_\alpha(y,z_{0:T},x)$ where the symbol vector encourages conditioning on a specific vector with the caveat that it must be learned by K-means clustering on the latents. In contrast, the decisions made for our model focus on generating anatomically realistic and data-driven reconstructions. This motivates our straightforward approach which avoids the symbol-vector coupling and intentionally adheres more strictly to the data, as is desirable for medical use cases and makes such an approach feasible for high dimensional data. Our approach utilizes $p_\alpha(z_{0:T},x)$ directly, making it more efficient and easier to train; indicatively, \cite{yu2022latent} applies their method on very low dimensional (D=2) synthetic data while our method can be easily applied to high dimensional image data. This high dimensionality is a key challenge which we explicitly sought to address to effectively and efficiently generate images.

\section{Method Details}
\subsection{Derivation of ELBO for LSD-EBM}\label{app:ELBO}
The objective function can be formulated with an evidence lower bound (ELBO) for $p_\theta(\mathbf{x})$, akin to the original \acp{vae} $(\theta := \alpha, \varphi, \beta)$:
\begin{equation}
\begin{aligned}
    \text{log} p_\theta(\mathbf{x}) &\geq 
       \mathbb{E}_{q_\varphi(\mathbf{z}_0|\mathbf{x})}\left[ \text{log} p_\beta(\mathbf{x}|\mathbf{z}_0)  \right] - D_{KL} \left(q_\varphi(\mathbf{z}_0|\mathbf{x}) || p_\alpha(\mathbf{z}) \right)  \\
     &= \mathbb{E}_{q_\varphi(\mathbf{z}_0|\mathbf{x})}\left[ \text{log} p_\beta(\mathbf{x}|\mathbf{z}_0) -  \text{log} q_\varphi(\mathbf{z}_0|\mathbf{x}) \right]  \\ 
     & \quad + \mathbb{E}_{q_\varphi(\mathbf{z}_0|\mathbf{x})}\left[  \text{log} p_\alpha(\mathbf{z}_0) \right] \\
     &=: \mathcal{L}(\alpha, \varphi, \beta).
\end{aligned}
\label{equ:lsd_elbo}
\end{equation}
The third term of $\mathcal{L}(\alpha, \varphi, \beta)$ is rewritten by Jensen's inequality at (I) as
\begin{equation}
\begin{aligned}
    &\mathbb{E}_{q_\varphi(\mathbf{z}_0|\mathbf{x})}\left[  \text{log} p_\alpha(\mathbf{z}_0) \right] \\
    &= \mathbb{E}_{q_\varphi(\mathbf{z}_0|\mathbf{x})}\left[\text{log} \int q(\mathbf{z}_{1:T}|\mathbf{z}_0)\frac{p_\alpha(\mathbf{z}_{0:T})}{q(\mathbf{z}_{1:T}|\mathbf{z}_0)}d\mathbf{z}_{1:T}  \right] \\
    &\overset{\mathrm{I}}{\geq} \mathbb{E}_{q_\varphi(\mathbf{z}_0|\mathbf{x})}\left[ \int q(\mathbf{z}_{1:T}|\mathbf{z}_0)\text{log}\frac{p_\alpha(\mathbf{z}_{0:T})}{q(\mathbf{z}_{1:T}|\mathbf{z}_0)}d\mathbf{z}_{1:T}  \right] \\
    &= \mathbb{E}_{q_\varphi(\mathbf{z}_0|\mathbf{x})q(\mathbf{z}_{1:T}|\mathbf{z}_0)}\left[ \text{log}\frac{p_\alpha(\mathbf{z}_{0:T})}{q(\mathbf{z}_{1:T}|\mathbf{z}_0)}  \right] \\
    &= \mathbb{E}_{q_\varphi(\mathbf{z}_0|\mathbf{x})q(\mathbf{z}_{1:T}|\mathbf{z}_0)} \left[\text{log} p(\mathbf{z}_T) + \sum_{t=0}^{T-1} \text{log}\frac{p_\alpha(\mathbf{z}_t | \mathbf{z}_{t+1})}{q(\mathbf{z}_{t+1}|\mathbf{z}_t)}   \right].
\end{aligned}
\label{equ:lsd_elbo_diffu}
\end{equation}

As $\mathbf{z}_T$ is in a standard Gaussian, $\text{log} p(\mathbf{z}_T)$ is a constant. Also, the conditional probabilities in \eqref{equ:lsd_elbo_diffu} is simplified to
\begin{equation}
\begin{aligned}
    & \text{log}p_\alpha(\mathbf{z}_t | \mathbf{z}_{t+1}) \\
    &= -\mathbb{E}_\alpha(\mathbf{z}_t, t) - \frac{1}{2\sigma_{t+1}^2}|| \mathbf{z}_{t+1} - \mathbf{z}_t ||^2 - \text{log} \tilde{Z}_{\alpha}(\mathbf{z}_{t+1}, t+1)\\
     &= -\mathbb{E}_\alpha(\mathbf{z}_t, t) - \frac{1}{2\sigma_{t+1}^2}|| \mathbf{z}_{t+1} -\tilde{\mathbf{z}}_t ||^2 \\ 
     & \quad -\mathbb{E}_{p_{\alpha}(\mathbf{z}_t | \mathbf{z}_{t+1})} \left[ -\mathbb{E}_\alpha(\mathbf{z}_t, t) - \frac{1}{2\sigma_{t+1}^2}|| \mathbf{z}_{t+1} - \mathbf{z}_t ||^2 \right]
\end{aligned}
\label{equ:lsd_elbo_cond}
\end{equation}

The objective function is finally
\begin{equation}
\begin{aligned}
    \mathcal{L}(\alpha, \varphi, \beta) &= \mathbb{E}_{q_\varphi(\mathbf{z}_0|\mathbf{x})}\left[ \text{log} p_\beta(\mathbf{x}|\mathbf{z}_0) -  \text{log} q_\varphi(\mathbf{z}_0|\mathbf{x}) \right]  \\
    & + \mathbb{E}_{q_\varphi(\mathbf{z}_0|\mathbf{x})q(\mathbf{z}_{1:T}|\mathbf{z}_0)}\sum_{t=0}^{T-1} \text{log}\frac{p_\alpha(\mathbf{z}_t | \mathbf{z}_{t+1})}{q(\mathbf{z}_{t+1}|\mathbf{z}_t)}, \\
\end{aligned}
\label{equ:lsd_elbo_objective}
\end{equation}
and the parameters $\alpha, \varphi, \beta$ are optimized by the gradient of $\mathcal{L}(\alpha, \varphi, \beta)$ as
\begin{equation}
\begin{aligned}
    \nabla_\theta\mathcal{L}(\alpha, \varphi, \beta) &= 
    \mathbb{E}_{q_\varphi(\mathbf{z}_0|\mathbf{x})}\left[ \nabla_\beta\text{log} p_\beta(\mathbf{x}|\mathbf{z}_0) -  \nabla_\varphi\text{log} q_\varphi(\mathbf{z}_0|\mathbf{x}) \right] +  \\& 
 \nabla_\alpha \mathbb{E}_{q_\varphi(\mathbf{z}_0|\mathbf{x})q(\mathbf{z}_{1:T}|\mathbf{z}_0)} \\
& \left[ \sum_{t=0}^{T-1} 
  -\mathrm{E}_{\alpha}(\mathbf{z}_t, t) - \mathbb{E}_{p_\alpha(\mathbf{z}_t | \mathbf{z}_{t+1})} \left[ -\mathrm{E}_{\alpha}(\mathbf{z}_t, t) \right]  \right]. \\
\end{aligned}
\label{equ:lsd_elbo_gradient}
\end{equation}

\subsection{Pseudo-algorithms for Latent Space Diffusion Energy-based Method}
\begin{algorithm}
\caption{Training of LSD-EBM}\label{alg:alg1}
\begin{algorithmic}
\STATE 
\STATE {\textbf{LOOP}}
\STATE \hspace{0.5cm} Select randomly $\mathbf{x}$
\STATE \hspace{0.5cm} $ \mathbf{z}_0 \sim q_\varphi(\mathbf{z}_0|\mathbf{x})$
\STATE \hspace{0.5cm} $ t \in{\{0, 1, ..., T-1\}}  $
\STATE \hspace{0.5cm} Compute $\mathbf{z}_t, \mathbf{z}_{t+1}$
\STATE \hspace{0.5cm} Get negative variable   $\tilde{\mathbf{z}}_t$ using \eqref{equ:mcmc}
\STATE \hspace{0.5cm} Compute reconstruction $\mathbf{x}^\prime \sim p_\beta(\mathbf{x}|\mathbf{z}_0)$
\STATE \hspace{0.5cm} Update $\beta, \varphi$ by the gradient
\STATE \hspace{0.5cm} $\quad \quad \nabla_\beta\text{log} p_\beta(\mathbf{x}^\prime|\mathbf{z}_0) -  \nabla_\varphi\text{log} q_\varphi(\mathbf{z}_0|\mathbf{x})$
\STATE \hspace{0.5cm} Update $\alpha$ by minimizing the energy loss 
\STATE \hspace{0.5cm} $\quad \quad -\mathrm{E}_{\alpha}(\mathbf{z}_t, t) - (-\mathrm{E}_{\alpha}(\tilde{\mathbf{z}}_t, t))$
\STATE {\textbf{UNTIL convergence}}
\end{algorithmic}
\label{alg1}
\end{algorithm}

\begin{algorithm}
\caption{Inference of LSD-EBM on 3D dataset}\label{alg:alg2}
\begin{algorithmic}
\STATE 
\STATE \textbf{Input:} $\mathbf{x}, T$
\STATE $ \mathbf{z}_0 \sim q_\varphi(\mathbf{z}_0|\mathbf{x})$
\STATE Compute $\mathbf{z}_T$
\STATE \textbf{for} $ t \in{\{T-1, ..., 0\}}  $ \textbf{do}
\STATE \hspace{0.5cm} Compute $\mathbf{z}_t$ given $\mathbf{z}_{t+1}$ using \eqref{equ:mcmc}
\STATE \textbf{End for}
\STATE Compute reconstruction $\mathbf{x}^\prime \sim p_\beta(\mathbf{x}|\mathbf{z}_0)$
\STATE \textbf{Return:} $\mathbf{x}^\prime$
\end{algorithmic}
\label{alg2}
\end{algorithm}

\begin{algorithm}
\caption{Inference of LSD-EBM on 2D dataset}\label{alg:alg3}
\begin{algorithmic}
\STATE 
\STATE $\mathbf{z}_T \sim \mathcal{N}(0, \mathbf{I})$
\STATE \textbf{for} $ t \in{\{T-1, ..., 0\}}  $ \textbf{do}
\STATE \hspace{0.5cm} Compute $\mathbf{z}_t$ given $\mathbf{z}_{t+1}$ using \eqref{equ:mcmc}
\STATE \textbf{End for}
\STATE Compute reconstruction $\mathbf{x}^\prime \sim p_\beta(\mathbf{x}|\mathbf{z}_0)$
\STATE \textbf{Return:} $\mathbf{x}^\prime$
\end{algorithmic}
\label{alg3}
\end{algorithm}

\section{Metrics} \label{app:metrics}
Let $S_A$ and $S_B$ be the two 3D reconstructions, $S_N(x, y, z) \in \{1, 0\}$ be the value of the pixel with the coordinates $x, y, z$ of segmentation $S_N$, and $|S_N(x, y, z)|$ is the total number of pixels (in our case, $128^3$). We use the following metrics~\cite{taha2015metrics}:

The Dice score between two segmentations $S_A$ and $S_B$ is defined as:
\begin{equation}
    \text{DICE}(S_A, S_B) = \frac{2 \times |S_A \cap S_B|}{|S_A| + |S_B|}
\end{equation}
where $|S_A \cap S_B|$ represents the volume of the intersection (i.e., the number of pixels or voxels that are positive in both $S_A$ and $S_B$), and $|S_A|$ and $|S_B|$ are the volumes (i.e., the total number of pixels) of segmentations $S_A$ and $S_B$, respectively.

Volumetric Similarity between two segmentations $S_A$ and $S_B$ is defined as:
\begin{equation}
    \text{VS}(S_A, S_B) = 1 - \frac{| |S_A| - |S_B| |}{|S_A| + |S_B|}.
\end{equation}

Specificity is defined as the proportion of true negatives (TN) out of the total number of actual negatives:
\begin{equation}
    \text{SPEC}(S_A, S_B) = \frac{\text{TN}}{\text{TN} + \text{FP}}
\end{equation}
where FP represents false positives.

Sensitivity, also known as recall or true positive rate, is defined as:
\begin{equation}
    \text{SEN}(S_A, S_B) = \frac{\text{TP}}{\text{TP} + \text{FN}}
\end{equation}
where TP represents true positives and FN represents false negatives.

Normalized Mutual Information (NMI) between $S_A$ and $S_B$ is defined as:
\begin{equation}
    \text{NMI}(S_A, S_B) = \frac{2 \times I(S_A; S_B)}{H(S_A) + H(S_B)}
\end{equation}
where $I(S_A; S_B)$ is the mutual information between $S_A$ and $S_B$, and $H(S_A)$ and $H(S_B)$ are the entropies of $S_A$ and $S_B$, respectively.

Cohen's Kappa (CK) is defined as:
\begin{equation}
    \text{CK}(S_A, S_B) = \frac{P_o - P_e}{1 - P_e}
\end{equation}
where $P_o$ is the relative observed agreement between $S_A$ and $S_B$, and $P_e$ is the hypothetical probability of chance agreement.

\section{Implementation Details} \label{app:implementation}

The \ac{vae}, \ac{lebm}, and \ac{lsdebm} are compared across results trained on different steps (\ac{mcmc} for the prior sampling in \ac{lebm}, and diffusion steps in \ac{lsdebm}), i.e., 2, 15, and 20 steps. The models \ac{vae}, \ac{lebm}, and \ac{lsdebm} are trained for 200 epochs with learning rates of $2 \times 10^{-5}$, $10^{-4}$, and $2 \times 10^{-5}$, and batch sizes of 4, 2, and 4 respectively. Training was performed on a NVIDIA A100 GPU with 40 GB memory. 

\section{Validation Experiments}\label{app:2D}

\begin{table}[htb]
\centering
\caption{2D datasets test FID scores}
\setlength{\tabcolsep}{3pt}
\begin{tabular}{|p{90pt}|p{55pt}|p{55pt}|p{55pt}|}
\hline
Dataset & EBM & LEBM & LSD-EBM \\
\hline
MNIST & 45.43  & 22.96  & \textbf{9.43}\\
FashionMNIST & 146.39 & 46.70 & \textbf{23.56}\\
CIFAR10 & 323.32 & \textbf{103.66}  & 108.71\\
CelebA & 360.05 & 43.32  & \textbf{27.89}\\
\hline
\end{tabular}
\label{tab:public_result}
\end{table}

We trained the \ac{vae}, \ac{lebm}, and \ac{lsdebm} on the standard 2D image datasets MNIST, FashionMNIST, CIFAR10, and CelebA and evaluated their performances using the FID score, shown in Table~\ref{tab:public_result}. The \ac{lsdebm} significantly outperforms its counterparts for MNIST, FashionMNIST and CelebA, and has comparable performance to the \ac{lebm} on the CIFAR10 dataset.  Critically, our model has increased variability in its generations as compared to the other methods, with further samples in Appendix~\ref{app:2D}. These results showcase the capability and generalizability of \ac{lsdebm} for image generation, and its application in the \cite{turella2021high} pipeline.

\begin{figure}[htb]
\centering
\begin{minipage}{1.5in}
    \centering
    \includegraphics[width=1.5in]{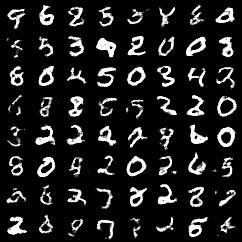}
    \label{ebm_mnist}
\end{minipage}
\begin{minipage}{1.5in}
    \centering
    \includegraphics[width=1.5in]{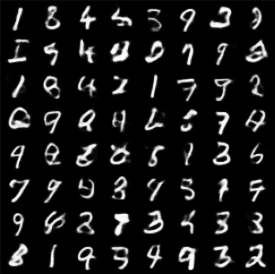}
    \label{lebm_mnist}
\end{minipage}
\begin{minipage}{1.5in}
    \centering
    \includegraphics[width=1.5in]{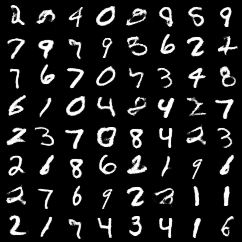}
    \label{lsd_ebm_mnist}
\end{minipage}
\caption{Examples of generated images by the models (EBM, LEBM, and LSD-EBM) trained on MNIST.}
\label{fig:mnist_gen}
\end{figure}

\begin{figure}[htb]
\centering
\begin{minipage}{1.5in}
    \centering
    \includegraphics[width=1.5in]{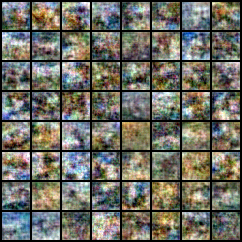}
    \label{ebm_cifar10}
\end{minipage}
\hfil
\begin{minipage}{1.5in}
    \centering
    \includegraphics[width=1.5in]{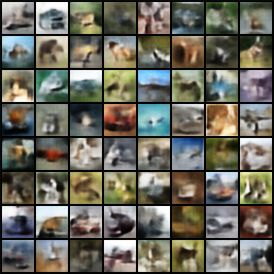}
    \label{lebm_cifar10}
\end{minipage}
\hfil
\begin{minipage}{1.5in}
    \centering
    \includegraphics[width=1.5in]{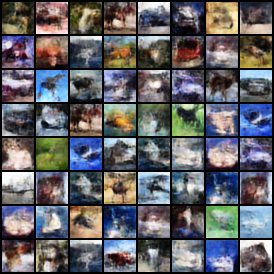}
    \label{lsd_ebm_cifar10}
\end{minipage}
\caption{Examples of generated images by the models (EBM, LEBM, and LSD-EBM) trained on CIFAR10.}
\label{fig:cifar10_gen}
\end{figure}

\begin{figure}[htb]
\centering
\begin{minipage}{1.5in}
    \centering
    \includegraphics[width=1.5in]{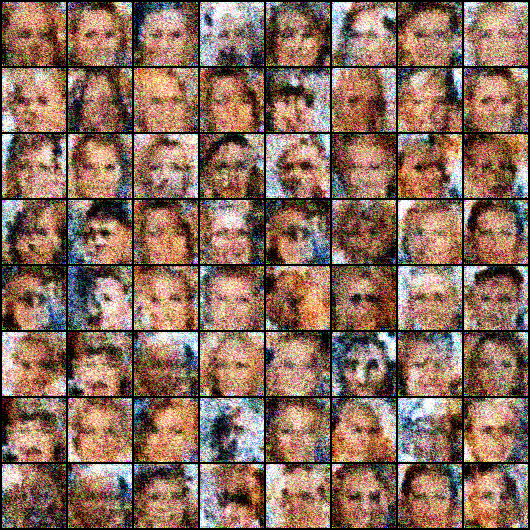}
    \label{ebm_celeba64}
\end{minipage}
\hfil
\begin{minipage}{1.5in}
    \centering
    \includegraphics[width=1.5in]{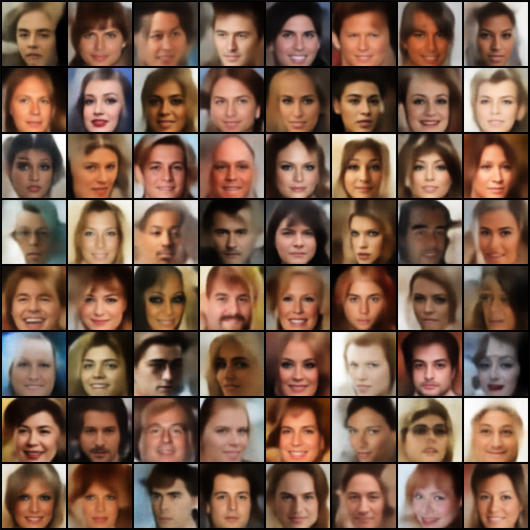}
    \label{lebm_celeba64}
\end{minipage}
\hfil
\begin{minipage}{1.5in}
    \centering
    \includegraphics[width=1.5in]{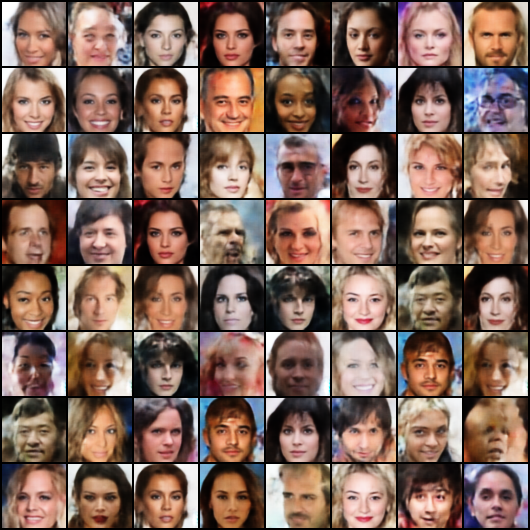}
    \label{lsd_ebm_celeba64}
\end{minipage}
\caption{Examples of generated images by the models (EBM, LEBM, and LSD-EBM) trained on CelebA.}
\label{fig:celeba64_gen}
\end{figure}


To test and compare our proposed method, we trained all models on the standard public image datasets: MNIST, CIFAR10, and CelebA.

The Fréchet Inception Distance (FID) score was used to assess the quality of images generated by models against a set of real images,

\begin{equation}
    \text{FID} = ||\mu_r - \mu_g||^2 + \text{Tr}(\Sigma_r + \Sigma_g - 2(\Sigma_r\Sigma_g)^{1/2})
\end{equation}

where $\mu_r$ and $\mu_g$ are the feature-wise mean vectors of the real and generated images, respectively, and $\Sigma_r$ and $\Sigma_g$ are the covariance matrices of the real and generated images, respectively.

A lower FID score indicates that the generated images are closer to the real images in terms of both content and style.

Our results are shown in Table.~\ref{tab:public_result}, and the corresponding generated images are visualized in Fig.~\ref{fig:mnist_gen}, Fig.~\ref{fig:cifar10_gen}, and Fig.~\ref{fig:celeba64_gen}. Our \ac{lsdebm} outperforms the other two methods on MNIST and CelebA, datasets that exhibit consistent similarities between images. On the CIFAR10 dataset, which is more challenging due to its random collection of images from different scenarios and lack of clear common characteristics within each category, \ac{lebm} performs better, though \ac{lsdebm} is a close second. These preliminary results collectively support the implementation of \ac{lsdebm} on the vertebrae segments as detailed in \cite{turella2021high}. 


\section{Time Comparison Across Methods} \label{app:time}
The training times for \ac{vae}, \ac{lebm}, and \ac{lsdebm} with 20 steps on the vertebrae dataset for 200 epochs are 12 hours, 33 hours, and 17 hours, respectively. The processing time of reconstruction of one vertebrae sample for \ac{vae}, \ac{lebm}, and \ac{lsdebm} are 0.039s, 0.65s, and 6.25s, respectively. The reconstruction time of \ac{lsdebm}, while slower, is well within acceptable bounds. Regarding computational efficiency, the processing of \ac{ddpm} with just two steps exceeds the 40 GB GPU memory limit, highlighting its inefficiency

\begin{table}[hbt!]
\centering
\begin{tabular}{|l|l|l|}
\hline
Model & Training Time (200 epochs) & Reconstruction Time (per sample) \\
\hline
\ac{vae} & 12h & 0.039s \\
\ac{lebm} & 33h & 0.65s \\
\ac{lsdebm} & 17h & 6.25s \\
\hline
\end{tabular}
\caption{Training and Reconstruction Times for Different Models on the Vertebrae Dataset}
\label{tab:model_times}
\end{table}

\section{Vertebrae Reconstruction Examples}\label{app:3D}

\begin{figure}[!ht]
\centering
    \includegraphics[width=4.8in]{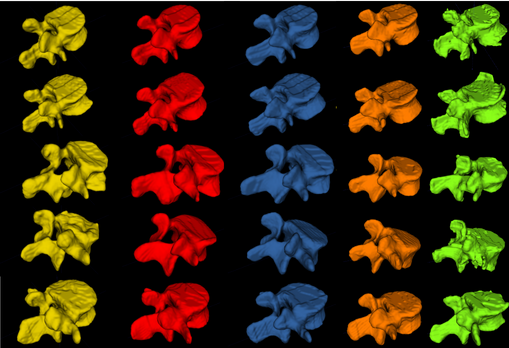}
    \caption{Reconstructions of the input (in yellow) using \ac{lebm}, \ac{vae}, and \ac{lsdebm} in order from left to right as compared to the ground truth (in green).}
\label{fig:extra_0}
\end{figure}

\begin{figure}[!ht]
\centering
    \includegraphics[width=4.8in]{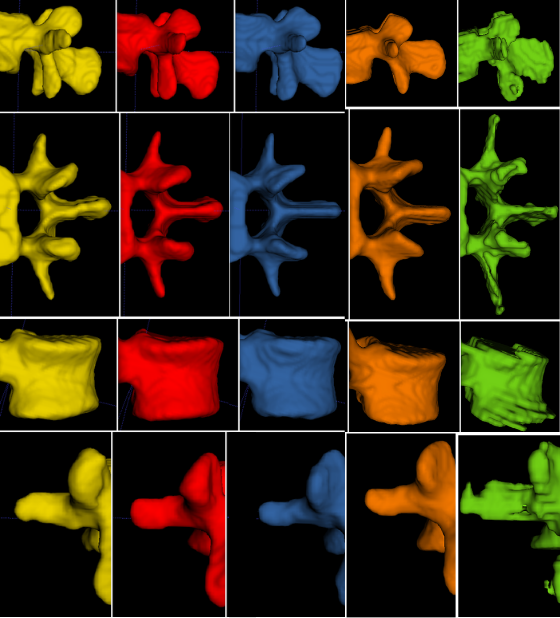}
    \caption{Reconstructions of the input (in yellow) using \ac{lebm}, \ac{vae}, and \ac{lsdebm} in order from left to right as compared to the ground truth (in green).}
\label{fig:extra_1}
\end{figure}

\begin{figure}[!ht]
\centering
    \includegraphics[width=4.8in]{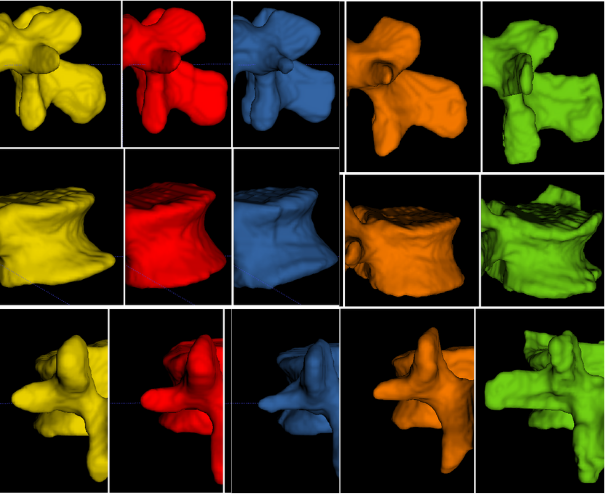}
    \caption{Reconstructions of the input (in yellow) using \ac{lebm}, \ac{vae}, and \ac{lsdebm} in order from left to right as compared to the ground truth (in green).}
\label{fig:extra_2}
\end{figure}

\begin{figure}[!ht]
\centering
    \includegraphics[width=4.8in]{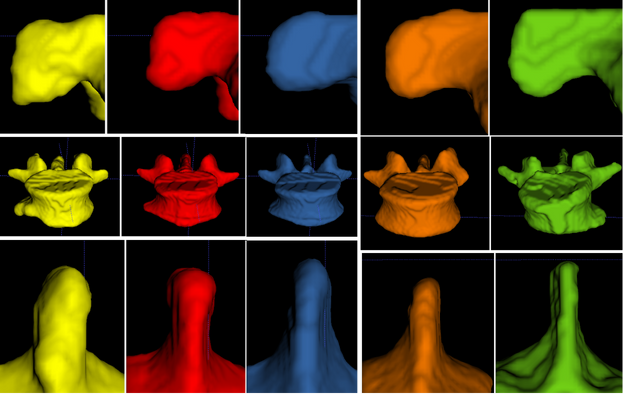}
    \caption{Reconstructions of the input (in yellow) using \ac{lebm}, \ac{vae}, and \ac{lsdebm} in order from left to right as compared to the ground truth (in green).}
\label{fig:extra_3}
\end{figure}

In Figs.~\ref{fig:extra_0},~\ref{fig:extra_1},~\ref{fig:extra_2},~\ref{fig:extra_3}, we show additional results comparing the input data, \ac{lebm}, \ac{vae}, and \ac{lsdebm}, and the ground truth high resolution model. We provide close-up details of regions of interest for closer comparison.

%





\end{document}